\documentclass[%
 reprint,
nofootinbib,
 amsmath,amssymb,
 aps,
]{revtex4-1}

\usepackage{graphicx}
\usepackage{dcolumn}
\usepackage{bm}
\usepackage{hyperref}
\usepackage{soul}

\usepackage{xcolor}
\newcommand{\red}[1]{}
\renewcommand{\red}[1]{{\color{red}{#1}}}

\newcommand{\blue}[1]{}
\renewcommand{\blue}[1]{{\color{blue}{#1}}}

\newcommand{\magenta}[1]{}
\renewcommand{\magenta}[1]{{\color{magenta}{#1}}}

\newcommand{\orange}[1]{}
\renewcommand{\orange}[1]{{\color{orange}{#1}}}

\newcommand{\rmd}{\mathrm{d}}

\newcommand{\ms}{m_\sigma}
\newcommand{\mr}{m_\rho}
\newcommand{\mf}{m_f}

\newcommand{\gs}{g_\sigma}
\newcommand{\gr}{g_\rho}
\newcommand{\gf}{g_f}

\newcommand{\GS}{\Gamma_\sigma}
\newcommand{\GR}{\Gamma_\rho}
\newcommand{\GF}{\Gamma_f}

\newcommand{\MeV}{\,\mathrm{MeV}}
\newcommand{\GeV}{\,\mathrm{GeV}}

\newcommand{\pt}{p\textsubscript{T}}
\newcommand{\twototwo}{2 $\leftrightarrow$ 2 }

\begin{document}


\title{Out-of-Equilibrium Photon Production in the Late Stages of Relativistic Heavy-Ion Collisions}

\author{Anna Sch\"afer$^{1,2,3}$}
\email{aschaefer@fias.uni-frankfurt.de}
\author{Oscar Garcia-Montero$^{2}$}
\author{Jean-Fran\c cois Paquet$^{4}$}
\author{Hannah Elfner$^{3,2,1,5}$}
\author{Charles Gale$^6$}

\address{$^1$ Frankfurt Institute for Advanced Studies (FIAS), Ruth-Moufang-Stra{\ss}e 1, 60438 Frankfurt am Main}
\address{$^2$ Institut f\"ur Theoretische Physik, Johann Wolfgang Goethe-Universit\"at, Max-von-Laue-Stra{\ss}e 1, 60438 Frankfurt am Main, Germany}
\address{$^3$ GSI Helmholtzzentrum f\"ur Schwerionenforschung, Planckstra{\ss}e 1, 64291 Darmstadt, Germany}
\address{$^4$ Department of Physics, Duke University, Durham, NC 27708, USA}
\address{$^5$ Helmholtz Research Academy Hesse for FAIR (HFHF), GSI Helmholtz Center,
Campus Frankfurt, Max-von-Laue-Stra{\ss}e 12, 60438 Frankfurt am Main, Germany}
\address{$^6$ Department of Physics, McGill University, 3600 University Street, Montreal, QC, H3A 2T8, Canada}

\date{\today}

\begin{abstract}
In this work, we assess the importance of non-equilibrium dynamics in the production of photons from the late stages of relativistic heavy-ion collisions.
The \pt{}-differential spectra and $v_2$ of photons from the late hadronic stage are computed within a non-equilibrium hadron transport approach, and compared to the results of a local equilibrium evolution using ideal relativistic hydrodynamics. It is found that non-equilibrium dynamics enhance the late-stage photon production at low \pt{} and decreases it at higher \pt{} compared to the estimate from hydrodynamics. This same comparison points to a significant increase in the momentum anisotropies of these photons due to non-equilibrium dynamics.
Once combined with photons produced above the particlization temperature in the hydrodynamics evolution, the differences between the two approaches appear modest in what concerns the \pt{} differential spectra, but are clearly noticeable at low \pt{} for the elliptic flow: non-equilibrium dynamics enhance the photon $v_2$ below \pt{} $\approx 1.4$ GeV.

\end{abstract}

\pacs{Valid PACS appear here}
\maketitle



\section{\label{sec:Intro}Introduction}

The collision of nuclei at relativistic energies has been shown to produce a plasma of quarks and gluons (QGP), an exotic state of matter which existed a few microseconds after the big bang \cite{[{See, for example, }][{, and references therein}]Shuryak:2014zxa}. A vibrant research program  to produce, study, and characterize the QGP is underway  at large accelerator facilities such as RHIC (Relativistic Heavy-Ion Collider) and at the LHC (Large Hadron Collider).
Several categories of experimental probes are invoked to study the quark-gluon plasma, each having their specific advantages and challenges. This work will focus on electromagnetic probes, which are not only {\it penetrating} as they have a long mean free path relative to the expanding medium, but can also be {\it soft} on a scale typical of hadronic physics. This combination makes photons and leptons unique and powerful tools in the collection of observables associated with the QGP \cite{[{See, for example, }][{, and references therein}]David:2019wpt}.

Photons are emitted at all stages of the collision -- from the very early moments to that of the very last radiative decay. As a consequence one has to create a distinction between "decay" and "direct" photons. The former is the largest source of photons, coming from hadronic decays, while the latter are produced dynamically via interactions of charged particles in the medium and elemental collisions and are obtained experimentally via the subtraction of the decay photons from the full spectrum. Direct photons, albeit being a small part of the total, contain the most information on the space-time evolution of the interacting medium.

The earliest sources of direct photons are those produced in hard partonic interactions at the impact of the nuclei. These so-called ``prompt photons'' completely dominate the direct (non-decay) photon signal at large transverse momentum \pt~\cite{Adler:2005ig,Adam:2015lda}, although their contribution to lower \pt{} demands progress in measurements and in theory.\footnote{Measurements are  scarce for direct photons with \pt~$\lesssim 2$ GeV produced in high-energy proton-proton collisions. On the other hand, collinear perturbative QCD calculations --- which can describe very well high-\pt{} photon measurements~\cite{Aurenche:2006vj} --- are known to break down at low \pt. Photon fragmentation functions are an additional source of uncertainty for lower \pt{} prompt photons.~\cite{Klasen:2014xfa,*Klasen:2013mga}}
Photons are also produced in out-of-equilibrium processes in the earlier stages of the plasma. These ``pre-equilibrium photons'' have been the object of significant attention over the past years~\cite{Garcia-Montero:2019kjk,*Churchill:2020uvk,*Garcia-Montero:2019vju,*Monnai:2019vup, *Gale:2020dum,*Gale:2020xlg,*Hauksson:2017udm,*Greif:2016jeb,*Oliva:2017pri,*Bhattacharya:2015ada}.
In the \pt{} range emphasized in this study, we expect pre-equilibrium photons to play a lesser role, as late-stage photons will have a larger $v_2$. However, their contribution should be quantified with an integrated hybrid approach; such studies have begun \cite{Gale:2021emg}.
Once the interacting medium approaches local equilibrium, its evolution can be described with hydrodynamics; and \textit{thermal} photons are produced as black-body radiation of the plasma. During this hydrodynamic expansion, the medium cools down and later reconfines into a gas of hadrons.
Eventually, the hydrodynamical description of this interacting medium has to be connected to hadronic transport, which provides a more realistic description for a dilute gas of hadrons and, importantly, allows for a dynamical freeze-out. During this last stage of the evolution, photons are produced continually from hadronic interactions and decays throughout this expansion and dilution of the medium. In this work, we focus on the direct photons emitted in binary mesonic interactions as well as bremsstrahlung from mesonic scatterings in a hadronic transport approach.

While hadronic measurements from the RHIC and the LHC can be described well with multistage, hybrid models of heavy-ion collisions (see e.g.~\cite{Schenke:2020mbo, Petersen:2014yqa} and references therein), photon data remain a challenge. The two main photonic observables are the \pt{} differential spectra and anisotropic flow, which have been measured at RHIC~\cite{Adare:2011zr,Adare:2014fwh, Adare:2015lcd} as well as at the LHC~\cite{Adam:2015lda, Acharya:2018bdy} for the midrapidity region. The simultaneous description of both these observables has been theoretically challenging~\cite{Paquet:2015lta, vanHees:2014ida, Kim:2016ylr,Shen:2013vja,Chatterjee:2011dw}, and has been dubbed the \textit{direct photon puzzle}, in the literature \cite{David:2019wpt}. A solid understanding of photon emission at later times is essential to understand RHIC and LHC measurements. The momentum anisotropy of \emph{direct} photons measured by PHENIX~\cite{Adare:2011zr,Adare:2015lcd} and ALICE~\cite{Acharya:2018bdy} is almost as large as that of pions. The momentum anisotropy of pions is understood to be imparted by the anisotropic flow velocity of the medium. This flow velocity anisotropy is small at earlier time, but large in the later stage of the collision where pions are produced. The large momentum anisotropy of direct photons therefore also demands an understanding of the electromagnetic emissivity during the later phases of the collision.

Up to now, photons produced by late stage hadronic interactions have in most cases been accounted for on a macroscopic basis, by combining a hydrodynamic description of the gas of interacting hadrons with thermal photon emission rates~\cite{Paquet:2015lta,vanHees:2011vb,Shen:2013vja,Chatterjee:2011dw,Chatterjee:2021bhz, Huovinen:2002im}.
Alternative previous efforts include photon production on an entirely microscopic basis ~\cite{Linnyk:2015rco}, as well as in a hybrid approach~\cite{Bauchle:2010ym}.\footnote{Note that, in contrast to the presented work, the latter includes hadronic photons from \twototwo scatterings relying on previous results for the underlying cross sections (see Ref. \cite{Kapusta:1992gv}). In addition, the results are limited to hadronic photon emission in \twototwo scatterings as contributions from bremsstrahlung processes are neglected.}
In this work, we provide a consistent calculation of hadron and photon production at the later stage of the collision employing such a hydrodynamics plus hadronic transport hybrid approach. In particular, the same hadronic model coupled to electromagnetism  that is used to calculate the thermal emission rates on the macroscopic side is used for the calculation of the corresponding cross-sections on the microscopic side. Therefore, we are able to assess the differences between local equilibrium and non-equilibrium photon emission from the late hadronic stage. Regarding the late stage photon emission below the particlization temperature (T = 150 MeV), we find notable difference between the local equilibrium and non-equilibrium approaches regarding the \pt{} spectrum and the differential $v_2$. Incorporating photons produced at earlier times, and thus higher temperatures ($T>150$~MeV), both approaches yield similar results for \pt{} spectra. The differential $v_2$ on the other hand show a clear enhancement at low \pt{} by virtue of non-equilibrium dynamics.

The hybrid model consists of (i) second-order relativistic hydrodynamics, as implemented in the code MUSIC \cite{Schenke:2010nt,Schenke:2010rr,Paquet:2015lta, MUSIC_link}, for the description of the near local equilibrium stage, followed by (ii) the hadronic transport code SMASH \cite{Weil:2016zrk, SMASH_doi, SMASH_github} in the  afterburner.
To assess the implications of non-equilibrium dynamics in the hadronic rescattering stage, we compare results for the hybrid and for the case where the hydrodynamical stage is extended well below the particlization temperature.

Our study is meant as a proof-of-concept study, which we have restricted to a simplified setting, omitting event-by-event fluctuations as well as viscous effects for the hydrodynamic evolution. Because of this, we refrain from a comparison to experimental data for the photon production. In addition, our results concern the mesonic interactions only and will have to be extended to interactions involving baryons to understand the interplay of the different stages, in particular when moving to heavy-ion reactions at lower beam energies, where finite net baryon densities are probed.

This work is organized as follows: In Section \ref{sec:Model} a detailed description of the hybrid approach is provided, as well as of the relevant contributions to photon production. The approach is corroborated by a decent agreement of hadronic \pt{} spectra and $v_2$ with experimental data. In section \ref{sec:Results} the results of photon production in a hybrid approach utilizing a non-equilibrium afterburner are presented. In the beginning, the focus is put on the spectra and anisotropies of photons produced during the afterburner stage. These results are compared to their hydrodynamical counterpart, extracted from thermal rates. In continuation, the full photon yields and anisotropies, including contributions from the earlier thermal QGP stage, are presented and discussed.
In Section \ref{sec:composition}, the composition of the late-stage photon \pt{} spectrum and elliptic flow is studied in greater detail while in Section \ref{sec:v2}, particular emphasis is put on the time evolution of the photon elliptic flow in the hadronic afterburner stage. Finally, a brief summary and outlook are given in Section \ref{sec:Outlook}.

%
%

\section{\label{sec:Model}Model Description}
\begin{figure}[t]
  \includegraphics[width=0.48\textwidth]{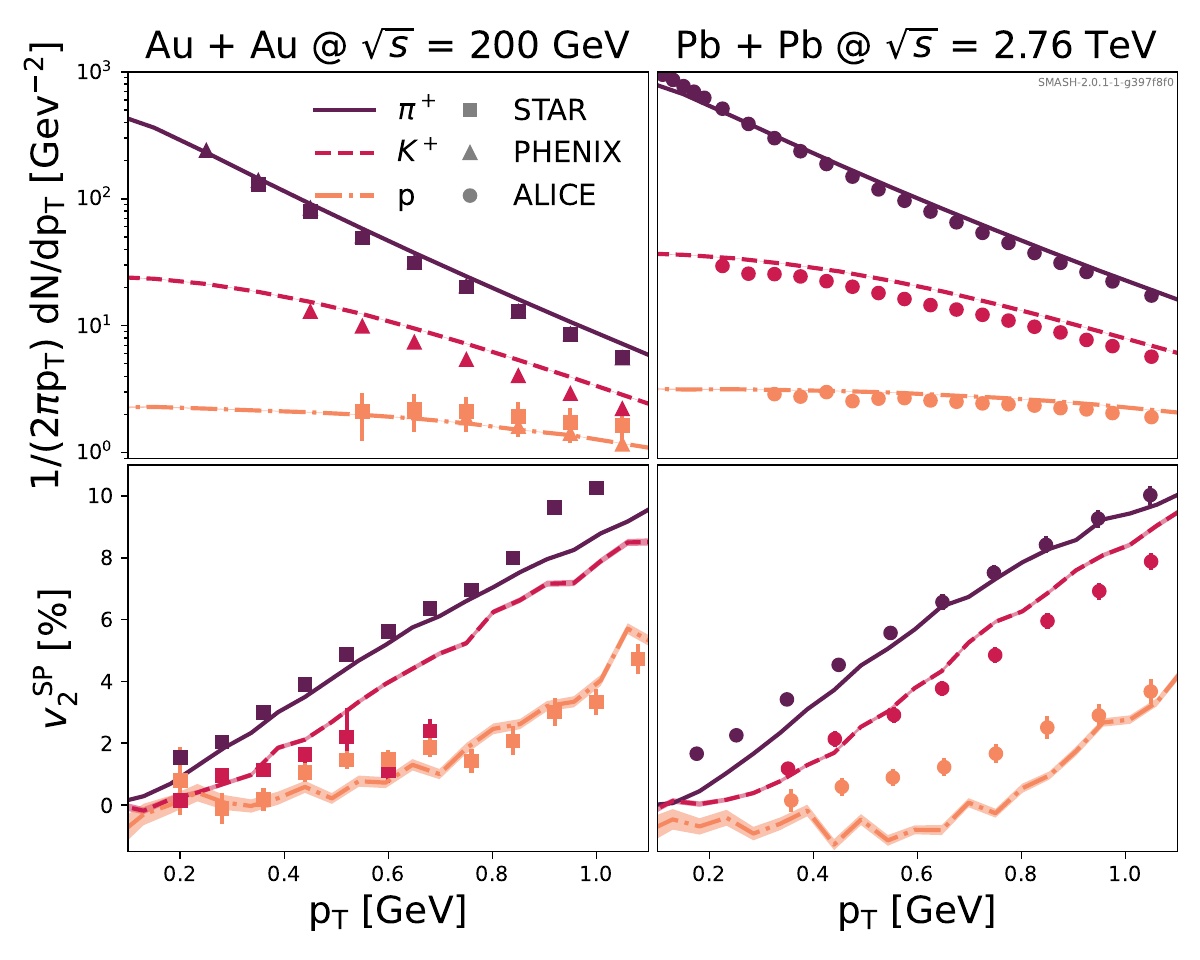}
  \caption{Pion, kaon and proton spectra from the MUSIC+SMASH hybrid in comparison to STAR \cite{Abelev:2006jr, Adams:2004bi}, PHENIX \cite{Adler:2003cb} and ALICE \cite{Abelev:2013vea, Abelev:2014pua} data in 10-20 \% most central collisions. The upper panel shows the \pt{} spectrum, and the lower panel the $v_2$. RHIC results are displayed on the left, LHC results on the right. The impact parameter used for the Trento event is b = 5 fm, to match approximately the data's 10-20 \% centrality bin.}
  \label{fig:hadron_spectra}
\end{figure}

The hybrid approach used in this work couples the relativistic hydrodynamics code MUSIC~\cite{Schenke:2010nt,Schenke:2010rr,Paquet:2015lta, MUSIC_link} to the hadronic transport approach SMASH \cite{Weil:2016zrk, SMASH_doi, SMASH_github}. In what follows, the initial conditions for the hydrodynamic evolution are obtained from the T\textsubscript{R}ENT\textsubscript{O} model~\cite{Moreland:2014oya}.
A single hydrodynamic event is simulated for each collision energy, and ideal 2+1D hydrodynamics is used.
The T\textsubscript{R}ENT\textsubscript{O} profile is used to initialize the transverse energy density of ideal hydrodynamics at $\tau=0.4$~fm/c. At this initial time, we set to zero the flow in the transverse directions.
The equation of state used in MUSIC matches lattice calculations to a hadron resonance gas with the same hadronic content as SMASH~\cite{Bazavov:2014pvz, Bernhard:2018hnz, eos_code}.
Particlization is enforced on a hypersurface of constant temperature, namely at $T = 150\MeV$.
A focus on ideal hydrodynamics allows us for a realistic first test case without the considerable uncertainties from the non-thermal corrections to the hadronic momentum distribution associated with viscous hydrodynamics (see, for example, Refs.~\cite{McNelis:2019auj,JETSCAPE:2020shq,Molnar:2020dvd} and references therein for a recent discussion).
With this combination of T\textsubscript{R}ENT\textsubscript{O} initial condition, ideal hydrodynamics and hadronic transport, hadronic observables are verified to show a fair agreement with midcentral spectra and $v_2$ measurements from RHIC and the LHC, as shown in Fig.~\ref{fig:hadron_spectra}. \\

When the medium is described with hydrodynamics, photons are emitted from the deconfined plasma phase as well as from the hadronic phase; there is in fact no sharp distinction between these two phases because of the cross-over nature of the transition (at zero net baryon density). Thermal photon emission is calculated by folding (i) thermal emission rates with (ii) the spacetime profile of temperature and flow velocity obtained from numerical hydrodynamic simulations.
For the thermal emission rate, we use the following prescription.
For temperatures above $180$~MeV, the electromagnetic emission rate used are those for a weakly-coupled quark-gluon plasma~\cite{Arnold:2001ms}, extrapolated to a fixed coupling of $g_s=2$; below $180$~MeV, hadronic thermal photon emission rates are used~\cite{Turbide:2003si,Liu:2007zzw,Heffernan:2014mla}.
The exact photon emission rates for temperatures in the QCD cross-over is still under investigation~\cite{Ce:2020tmx,Jackson:2019yao,Gale:2014dfa}.
To approximate the transition from partonic to hadronic rates in this temperature regime, we choose a transition temperature of $T=180$~MeV, as done previously in the literature \cite{Paquet:2015lta}.

For the purpose of the presented work, it is important that the hadronic thermal photon rate used with the hydrodynamics profiles includes only emission channels that are implemented in SMASH, which are:

\vspace{+0.2cm}
\begin{itemize}
    \item 2 $\leftrightarrow$ 2 scatterings: $\pi \rho \to \pi \gamma$  \cite{Turbide:2003si}
    \item Bremsstrahlung: $\pi \pi \to \pi \pi \gamma$ \cite{Liu:2007zzw,Heffernan:2014mla}
\end{itemize}
\vspace{+0.2cm}
These channels constitute a subset of the total hadronic photon production processes. However, owing to a large coupling and to the abundance of pions and $\rho$ mesons, they provide the leading contributions for photon emission in a hadronic medium \cite{SimonT,Gale:2003iz,Gale:2003iz}.

Unlike hydrodynamics, photon production in SMASH does not rely on thermal rates, but rather on individual photons produced by microscopic scatterings in the transport approach. This means that the input needed is explicit cross-sections for the scatterings involving photons and hadrons. For details about their derivation and the numerical implementation, the reader is referred to Appendix \ref{app:SMASH_brems} and Ref. \cite{Schafer:2019edr}. Note that in the past, the soft photon approximation has been applied for the microscopic production of hadronic bremsstrahlung photons \cite{Low:1958sn}. In our work, we go beyond this approximation, because there are regions in the phase space of the sampled photons for which the soft photon approximation is not applicable. For example, for incoming pion energies between the mediator masses (0.7-1.2 GeV), where photon production is enhanced by resonant peaks, the soft-photon approximation breaks down already at transverse photon momenta larger than 100 MeV.

\begin{figure}[t]
  \includegraphics[width=0.39\textwidth]{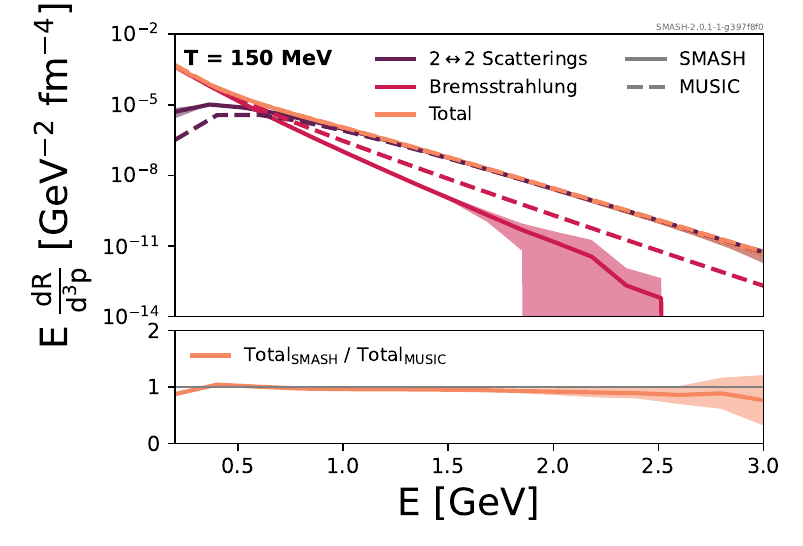}
  \caption{Comparison of the thermal photon rates at T = 150 MeV as used in MUSIC (dashed lines) to the photon rate extracted from SMASH (solid lines) in an infinite matter setup, according to their process origins. The lower panel shows the ratio of the total thermal photon rates from SMASH to those used in MUSIC.}
  \label{fig:thermal_rate}
\end{figure}

We verify our calculation by extracting thermal photon rates from \twototwo and bremsstrahlung processes from SMASH in an infinite matter setup at a temperature of T = 150 MeV, and comparing them to the corresponding thermal emission rates from Refs.~\cite{Turbide:2003si,Liu:2007zzw,Heffernan:2014mla}. The results are displayed in Fig.~\ref{fig:thermal_rate}, where full lines denote the SMASH results, and dashed lines the thermal rates used with hydrodynamics.
The agreement is good, although not perfect; these differences are expected for the following reasons.
First, the \twototwo rates from Ref.~\cite{Turbide:2003si} rely on a stable $\rho$ meson with zero width, although in reality $\Gamma_\rho \approx$  0.149 GeV. As it is comparably straightforward to account for this width in SMASH, we have decided to make use of this feature for the sake of a more realistic description rather than a perfectly equivalent comparison. The difference in the equilibrium photon rates introduced by choosing $\Gamma_\rho$ = 0.149 GeV can be quantified to result in an enhancement up to a factor of $10^3$ for very low photon energies -- where the vector meson is largely off-shell -- but becomes negligible above E$_\gamma$ $\approx$~1~GeV. The deviation in the bremsstrahlung photon rate is related to the choice of parameters for the computation of the corresponding cross sections. The masses, widths and couplings are chosen such that they match the SMASH degrees of freedom (see Appendix~\ref{app:SMASH_brems} for further details). These parameters are however slightly different from those used in Refs.~\cite{Liu:2007zzw, Heffernan:2014mla}. Deviations, especially at higher photon energies, are therefore expected.
Overall, the total photon rates of the two approaches are in good agreement, as displayed in the lower panel of Fig.~\ref{fig:thermal_rate}. This is because in the low energy regime, where the differences in the $\rho$ treatment are most striking, the thermal photon rate is dominated by bremsstrahlung photons. In the higher energy regime, where the mismatch of the bremsstrahlung parameters becomes of relevance, the bremsstrahlung contribution is subleading and the thermal photon rate is dominated by photons from \twototwo scatterings.

To assess the production of photons in the late stage of heavy-ion collisions, two different approaches are compared:
\vspace{+0.2cm}
\begin{itemize}
    \item[A:] T\textsubscript{R}ENT\textsubscript{O} + Hydrodynamics ($T>150$~MeV) + Hadronic transport
    \item[B:] T\textsubscript{R}ENT\textsubscript{O} + Hydrodynamics ($T>150$~MeV) + \mbox{Hydrodynamics (150~MeV $>T>$120~MeV)}
\end{itemize}
\vspace{+0.2cm}
where ``setup A'' corresponds to the transport (non-equilibrium) description of the late hadronic stage\footnote{To obtain the afterburner results in ``setup A'', SMASH version \texttt{SMASH-2.0.1-1-g397f8f0} is applied and the results are averaged over 40,000 events. Ten ``fractional photons'' are used to properly sample the photon angular distributions (see Appendix~\ref{app:SMASH_brems} for an explanation of fractional photons).
}
and ``setup B'' to the approximation of this stage using ideal hydrodynamics. Identical initial conditions are used for both setups.\\

%
%

\section{\label{sec:Results}Results}
The hybrid approach detailed in the last section is used to calculate photon production in heavy-ion collisions at RHIC and at the LHC: Au+Au collisions at $\sqrt{\mathrm{s_{NN}}}$ = 200 GeV, and Pb+Pb collisions at $\sqrt{\mathrm{s_{NN}}}$ = 2.76 TeV, respectively. In both cases an impact parameter of b = 5 fm was chosen as a proxy for a mid-central collision (see Fig.~\ref{fig:hadron_spectra}).

\begin{figure}[t]
  \includegraphics[width=0.49\textwidth]{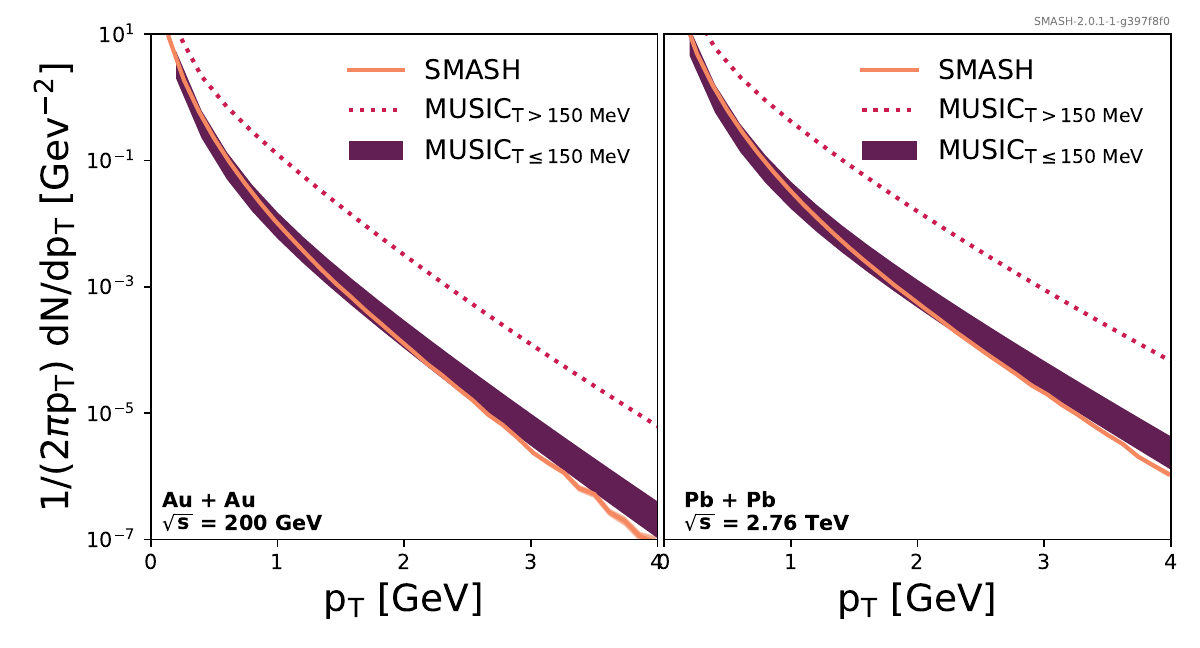}
  \caption{p$_\text{T}$ spectra of photons from the late hadronic stage in Au+Au collisions at $\sqrt{\mathrm{s}_\mathrm{NN}}$ = 200 GeV (left) and Pb+Pb collisions at $\sqrt{\mathrm{s}_\mathrm{NN}}$ = 2.76 TeV (right). Results of a non-equilibrium treatment in the afterburner are displayed by lines, while bands correspond to the results of hydrodynamics ($T<150$~MeV) and thermal rates. For the latter, the lower and upper limits are provided by radiating photons down to $T=140$~MeV and down to $T=120$~MeV, respectively. The dotted line denotes the contribution of photons originating from hydrodynamics above $T=150$~MeV. The impact parameter of the event is $b=5$~fm, corresponding approximately to a centrality bin of 10-20\%.}
  \label{fig:pTspectra}
\end{figure}

Fig.~\ref{fig:pTspectra} shows the total \pt{} spectrum of photons emitted in the late hadronic stage at RHIC energies on the left and LHC energies on the right. The dotted line corresponds to photon emission from hydrodynamics above $T=150$~MeV. Photon emission from SMASH is denoted with solid lines, while bands are used for photons calculated with thermal rates and hydrodynamics below $T=150$~MeV. The reader is reminded that the former corresponds to a non-equilibrium description of the interacting hadrons; the latter assumes that the evolution of the hadrons can be approximated as that of an ideal fluid, and their photon emission with thermal rates. The bands denote the uncertainty that lies in choosing the region of spacetime at which thermal electromagnetic radiation is produced, as determined by contours of uniform temperature.
The lower limit of the band corresponds to photon emission between $T= 150$~MeV (the particlization temperature used in this work) and $T=140$~MeV; the upper limit of this same band is for photons radiated between $T=150$~MeV and $T = 120$~MeV. Note that these small changes in temperature correspond to a significant increase in spacetime volume of photon emission, given the rate of cooling of the plasma at late times.\footnote{A rough approximation can be calculated with the conformal ideal Bjorken hydrodynamics formula $\tau_2=\tau_1 (T_2/T_1)^{3}$. According to this approximation, it takes $(140/120)^3\sim 1.6$ times longer to reach T = 120 MeV than it took to reach T = 140 MeV, a substantial increase in the spacetime volume. This is an approximation, given that the medium cools down more rapidly in $2+1$D than in $0+1$D, which is partly compensated by the slower cool down rate --- we used the speed of sound $c_s^{2}=1/3$ instead of the smaller $c_s^{2}\sim 1/4-1/5$ for QCD at temperatures of $100$ to $150$~MeV.}

For \pt~$\lesssim 2$~GeV, the photon spectrum obtained in the late stage of the non-equilibrium (``setup A'') lies entirely within the band determined in the late stage of the local equilibrium (``setup B'').
In general, this is the most relevant range of \pt{} for photons produced at lower energy densities: higher \pt{} photons tend to be dominated by photons produced at higher temperature, or by prompt photons.
Overall, the spectra of photons from the non-equilibrium approach is softer, however.
This is visible over the whole \pt{} range, although it is more evident for \pt{}~$\gtrsim 2$~GeV, where the non-equilibrium results fall outside of the band calculated with hydrodynamics and thermal rates.
As discussed in the previous section,
there are differences between the thermal photon emission rates calculated with SMASH and the one used in combination with hydrodynamics. However, this difference is very small in the combined bremsstrahlung and $\pi \rho \to \pi \gamma$ rates: too small to explain the softening of the spectra observed in Fig.~\ref{fig:pTspectra}.
We thus attribute this softening to non-equilibrium effects.

\begin{figure}[t]
  \includegraphics[width=0.48\textwidth]{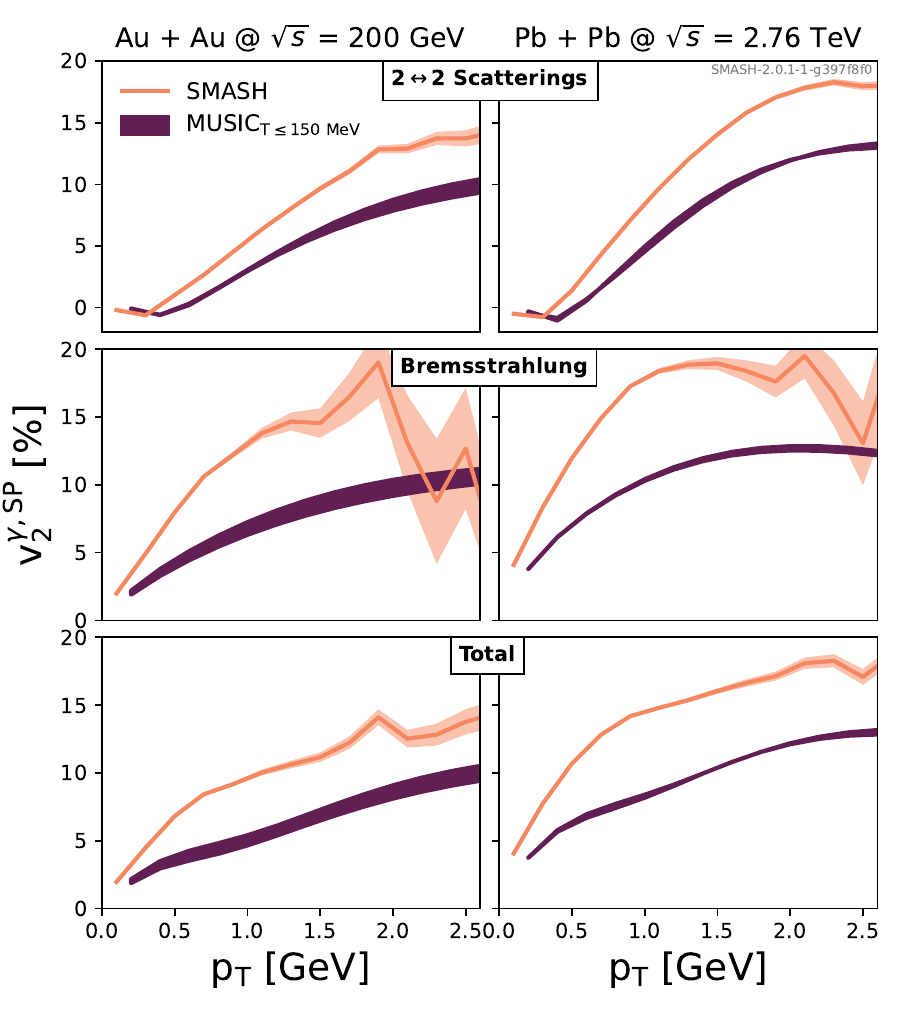}
  \caption{$v_2$ of photons from the late hadronic stage in Au+Au collisions at $\sqrt{\mathrm{s}_\mathrm{NN}}$ = 200 GeV (left) and Pb+Pb collisions at $\sqrt{\mathrm{s}_\mathrm{NN}}$ = 2.76 TeV (right) according to their production processes. Results of a non-equilibrium treatment in the afterburner are displayed by lines, while bands correspond to the results of the low-temperature hydrodynamic description. For the latter, the lower and upper limits are provided by radiating photons from $T=150$~MeV down to $T=140$~MeV and down to $T=120$~MeV, respectively. Bands accompanying the SMASH curves denote the statistical uncertainty therein. The impact parameter of the event is $b=5$~fm, corresponding approximately to a centrality bin of 10-20\%.}
  \label{fig:v2spectra}
\end{figure}

The photon $v_2$ is calculated with the scalar product method (see Refs.~\cite{Paquet:2015lta,Kim:2016ylr}); it is shown in  Fig.~\ref{fig:v2spectra}, again only for late-stage photons. Results for RHIC collisions can be found on the left, and LHC's on the right. The dark bands correspond to the hydrodynamic description of the late stage hadronic medium (``setup B''), and the lines surrounded by lighter (statistical error) bands to the non-equilibrium hadronic transport treatment (``setup A''). The top panels show the elliptic flow of photons originating from \twototwo scatterings, the middle panel the elliptic flow of photons from bremsstrahlung processes, and the bottom panel the combined results, properly weighted by their spectra. For \twototwo scatterings at low \pt, both SMASH and the hydrodynamic approach have a small $v_2$. This is understood to be a consequence of the weak dependence on energy of the \twototwo rate at low energy: softer rates lead to smaller $v_2$'s than harder rates. At higher \pt{}, the $v_2$ from SMASH is larger than that obtained with ideal hydrodynamics.
For bremsstrahlung photons, the $v_2$ from SMASH is significantly larger than the $v_2$ obtained with hydrodynamics and thermal rates. While Fig.~\ref{fig:thermal_rate} did show that the bremsstrahlung thermal rate in SMASH is somewhat softer than the thermal rates used with hydrodynamics, we verified that this is a $\sim 10\%$ effect that does not explain the much larger $v_2$ obtained in SMASH.
The combined $v_2$ from bremsstrahlung and \twototwo scatterings is also found to be significantly higher from SMASH than from hydrodynamics, consistent with the larger $v_2$ observed for the individual channels. We attribute the clear enhancement of photon $v_2$ in the case of SMASH to non-equilibrium effects.  It can be quantified that the total $v_2$ from the SMASH hadronic transport is, relative to that obtained from hydrodynamics, enhanced by a factor of 1-2 at RHIC as well as LHC in the presented p$_\mathrm{T}$ range.

\begin{figure}[t]
  \includegraphics[width=0.48\textwidth]{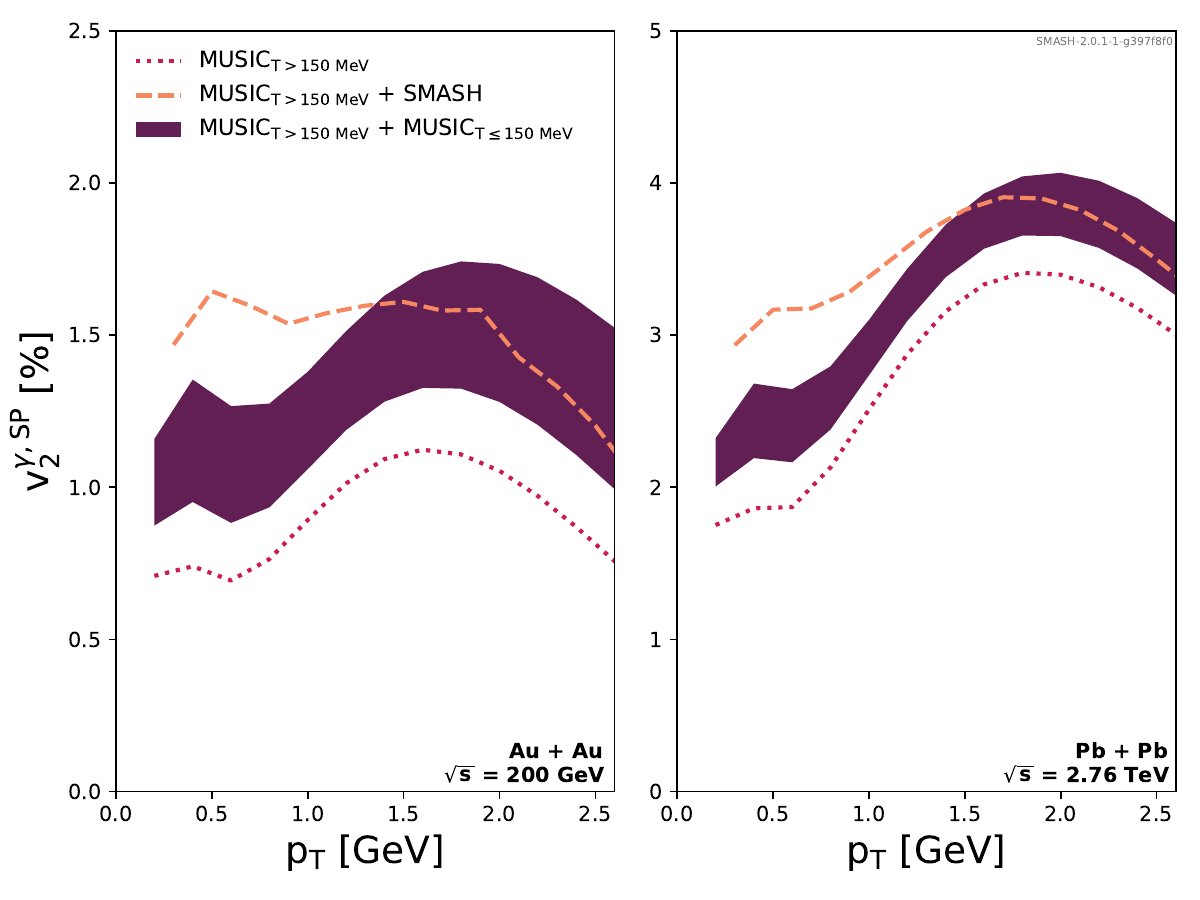}
  \caption{Combined $v_2$ of photons from the QGP and the hadronic afterburner stage in Au+Au collisions at $\sqrt{\mathrm{s}_\mathrm{NN}}$ = 200 GeV (left) and Pb+Pb collisions at $\sqrt{\mathrm{s}_\mathrm{NN}}$ = 2.76 TeV (right). Results applying a non-equilibrium afterburner are denoted with dashed lines, the low-temperature hydrodynamic description with bands, and the thermal photons from $T>150$~MeV with dotted lines. Note that the lower and upper limits of bands are obtained by radiating photons from $T=150$~MeV down to $T=140$~MeV and down to $T=120$~MeV, respectively. The impact parameter of the event is $b=5$~fm, corresponding approximately to a centrality bin of 10-20\%.
  }
  \label{fig:v2spectra_sum}
\end{figure}

Summarizing the findings from Figs.~\ref{fig:pTspectra} and~\ref{fig:v2spectra}, it can clearly be stated that non-equilibrium dynamics in the hadronic afterburner have notable implications for the \pt{} spectra  of late-stage photons, as well as their $v_2$.
While the \pt{} spectra exhibit a small softening in the transport case, compared to the hydrodynamic approach, $v_2$ is enhanced by up to a factor of 2 in the presented \pt{} range.\\
\begin{figure}[t]
  \includegraphics[width=0.48\textwidth]{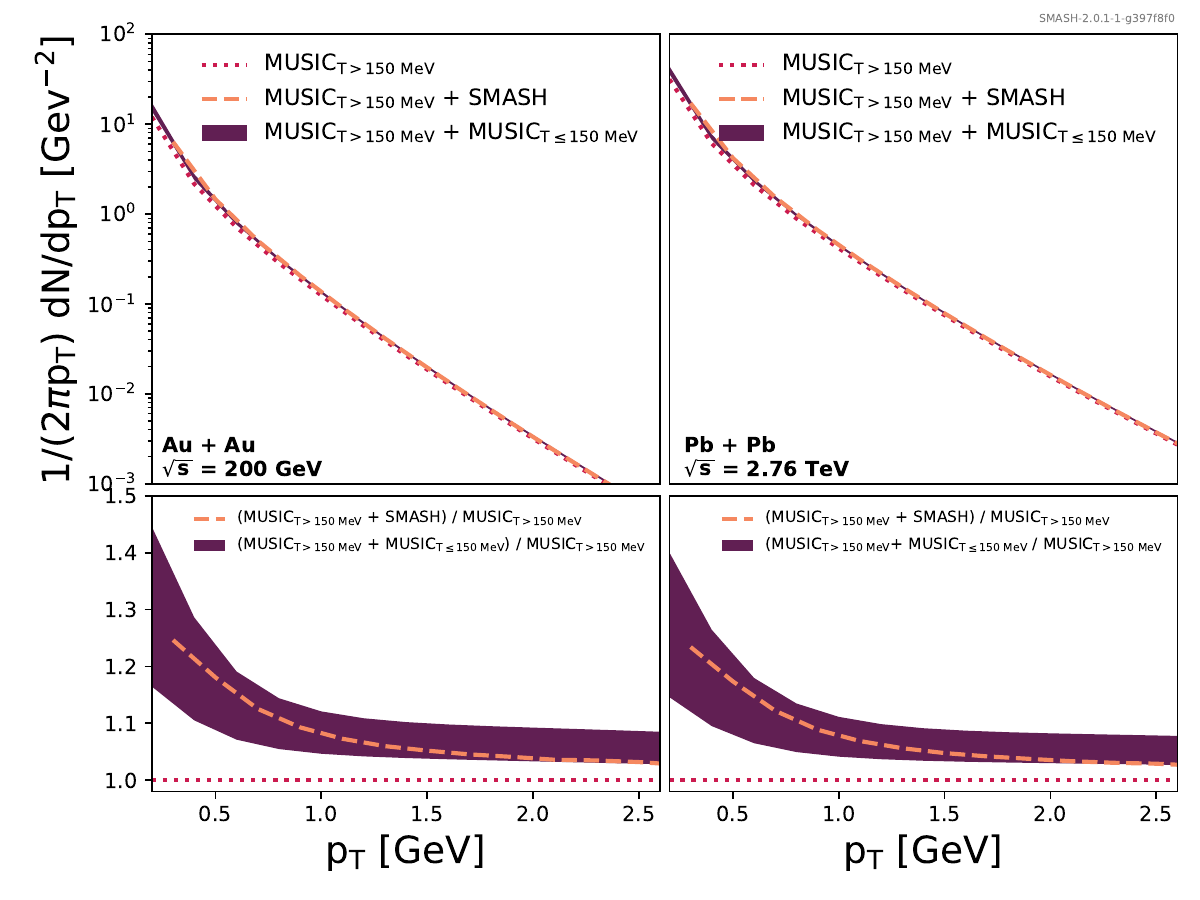}
  \caption{Combined p$_\text{T}$ spectra of photons produced at T $>$ 150 MeV and in the hadronic afterburner stage in Au+Au collisions at $\sqrt{\mathrm{s}_\mathrm{NN}}$ = 200 GeV (left) and Pb+Pb collisions at $\sqrt{\mathrm{s}_\mathrm{NN}}$ = 2.76 TeV (right). Results applying a non-equilibrium afterburner are denoted with dashed lines, the low-temperature hydrodynamic description with bands and the high-temperature hydrodynamic description with dotted lines. The lower panel shows the ratio, normalized to the high-temperature hydrodynamic description. Note that the lower and upper limits of the low-temperature hydrodynamic description are obtained by radiating photons from $T=150$~MeV down to $T=140$~MeV and down to $T=120$~MeV, respectively. The impact parameter of the event is $b=5$~fm, corresponding approximately to a centrality bin of 10-20\%.
  }
  \label{fig:pTspectra_sum}
\end{figure}
It is important to assess how significant these differences remain once combined with the large number of thermal photons produced in the earlier phase of the collision. For this, the full in-medium picture is obtained by combining hadronic photons from both approaches with the thermal radiation emitted with $T>150$~MeV. In Figs. \ref{fig:pTspectra_sum} and \ref{fig:v2spectra_sum}, the pink dotted lines show the contribution with $T>150$~MeV as obtained by hydrodynamics. The full photon spectra from the non-equilibrium afterburner (``setup A'') are denoted with orange dashed lines; those estimated with hydrodynamics and thermal rates at late times (``setup B'') with wide bands. For readability, Fig.~\ref{fig:pTspectra_sum} also contains a ratio plot in the lower panel, where the hydrodynamics as well as the non-equilibrium contributions combined with the $T>150$~MeV contribution are normalized to the $T>150$~MeV contribution. \\
\indent It is found that, at RHIC and LHC energies, the photon \pt{} spectra in Fig.~\ref{fig:pTspectra_sum} obtained in the non-equilibrium ``setup A'' lie within the bands provided by the local-equilibrium ``setup B''. The effect of softer \pt{} spectra in ``setup A'', as observed in Fig.~\ref{fig:pTspectra}, are visible in the ratios, where at lower \pt{} the ``MUSIC + SMASH'' line lies at the upper end of the ``MUSIC + MUSIC'' band, and at its lower end for higher \pt{}. Still, a proper non-equilibrium treatment in the late stages has only a minor impact on the final photon \pt{} spectra, once combined with contributions from hydrodynamics at $T>150$~MeV.\\
In the case of the differential $v_2$ however, some of the differences
observed in Fig.~\ref{fig:v2spectra} for the pure afterburner stage between the non-equilibrium and the local-equilibrium setups are still apparent once combined with photons produced above $T=150$~MeV. In Fig.~\ref{fig:v2spectra_sum} it is shown that the results obtained with the non-equilibrium ``setup A'' lie within the bands provided by the local-equilibrium ``setup B'' for \pt{} $\gtrsim 1.4$ GeV; however, at lower \pt{}, the difference between the two approaches is clearly visible. This observation holds for RHIC as well as for LHC. The fact that the differences found in Fig.~\ref{fig:v2spectra} are propagated only to low \pt{} can be explained as follows:
at high \pt{}, the vast majority of photons stem from hydrodynamics at $T>150$~MeV, exceeding contributions from $T\leq150$~MeV by multiple orders of magnitude (c.f. Fig.~\ref{fig:pTspectra}). As a result, the combined $v_2$ is mostly dominated by contributions from hydrodynamics at $T>150$~MeV, thus drowning the signal coming from the later stages of the evolution.
Moving to lower \pt{}, the relative contribution of photons produced in the afterburner stage is, although not dominant, significantly higher (c.f. Fig.~\ref{fig:pTspectra}).
Consequently, the $v_2$ carried by these photons isn't as diluted at low \pt{} than at high \pt{}.
Furthermore, since the $v_2$ of photons produced in the late stage of the non-equilibrium ``setup A'' largely exceeds that of photons produced with hydrodynamics at $T>150$~MeV, this excess is sufficient to significantly increase the resulting $v_2$.
Quantitatively, the $v_2$ determined within the non-equilibrium ``setup A'' is enhanced by up to 30\% at RHIC and up to 20\% at LHC as compared to the local-equilibrium ``setup B'' if photons are emitted down to $T=120$ MeV.
Proper non-equilibrium dynamics in the late stages are thus of major importance for photon $v_2$ at low \pt{}. We find that the application of an equilibrium treatment within hydrodynamics is not sufficient to properly capture the underlying dynamics that lead to an enhancement of $v_2$ out of equilibrium. \\

\subsection{Composition of the late-stage photons}
\label{sec:composition}
\begin{figure}[t]
	\includegraphics[width=0.48\textwidth]{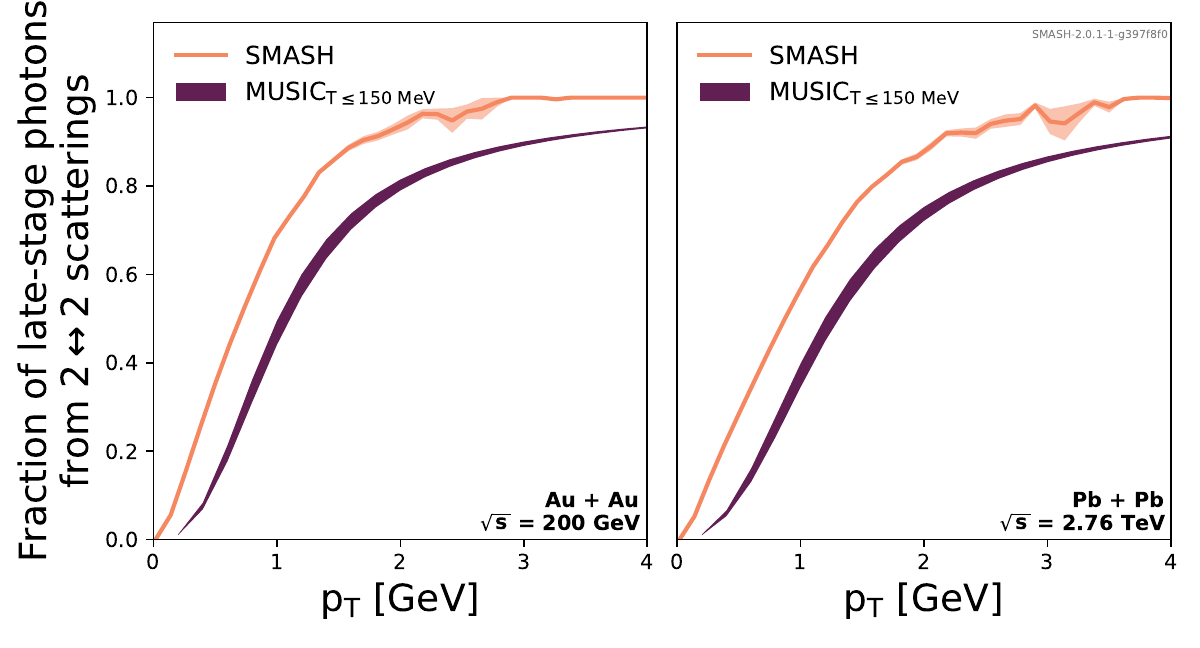}
	\caption{Fraction of late-stage photons originating from \twototwo processes, compared to the sum of \twototwo and bremsstrahlung, as a function of \pt{}. Left panel is for Au+Au collisions at $\sqrt{\mathrm{s}_\mathrm{NN}}$ = 200 GeV and right panel for Pb+Pb collisions at $\sqrt{\mathrm{s}_\mathrm{NN}}$ = 2.76 TeV. Results of a non-equilibrium treatment in the afterburner are displayed by the lighter lines, while the darker bands correspond to the results of hydrodynamics ($T<150$~MeV) and thermal rates. For the latter, the lower and upper limits of the bands are provided by radiating photons from $T=150$~MeV down to $T=140$~MeV and down to $T=120$~MeV, respectively. Bands accompanying the SMASH curves denote the statistical uncertainty therein.}
	\label{fig:ratio_spectra}
\end{figure}
In this section, the channel of production of photons in the late afterburner stage is detailed and subsequently related to the resulting $v_2$ coefficients. The relative contribution of photons from \twototwo scatterings as a function of transverse momentum is displayed in Fig.~\ref{fig:ratio_spectra} for collisions at RHIC on the left and LHC on the right. Results from the non-equilibrium afterburner (``setup A'') are shown with orange lines, those from the hydrodynamical description (``setup B'') with purple bands. The reader is reminded that, in the presented framework, photons can only be produced in \twototwo scatterings or via bremsstrahlung. The remaining contribution in Fig.~\ref{fig:ratio_spectra} thus corresponds to that of bremsstrahlung photons.
It can be observed that at low \pt{}, late-stage photons are produced dominantly by bremsstrahlung processes, while at high \pt{}, \twototwo scatterings dominate. This is expected from the value of the thermal rates seen in Fig.~\ref{fig:thermal_rate}. It is also consistent with the relatively low scattering energies of pions in the afterburner: bremsstrahlung photons with \pt{} $>$ 2.5 GeV are rarely produced in the transport evolution.
Furthermore, the transition point between bremsstrahlung and \twototwo occurs at a smaller \pt{} for SMASH than for the hydrodynamics, which is again explained by the difference in the rates shown in Fig.~\ref{fig:thermal_rate}.\\

\begin{figure}[t]
  \includegraphics[width=0.48\textwidth]{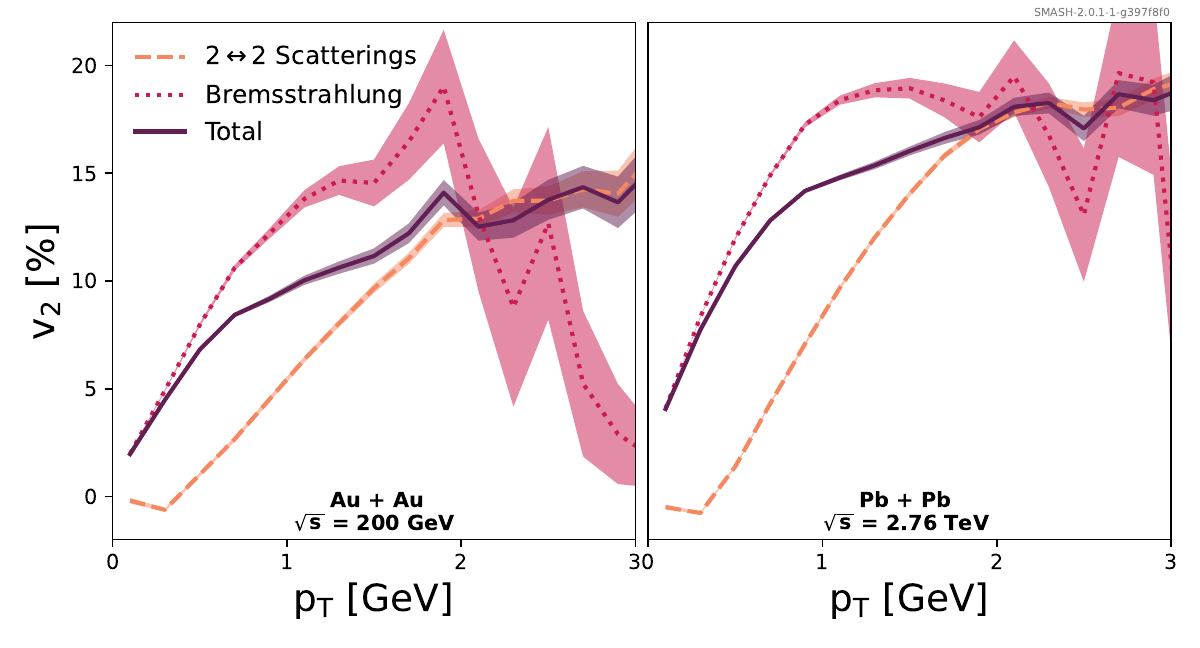}
  \caption{$v_2$ of photons from the late hadronic stage in Au+Au collisions at $\sqrt{\mathrm{s}_\mathrm{NN}}$ = 200 GeV (left) and Pb+Pb collisions at $\sqrt{\mathrm{s}_\mathrm{NN}}$ = 2.76 TeV (right) as obtained from the non-equilibrium SMASH afterburner. Dashed lines correspond to photons from \twototwo scatterings, dotted lines to photons from bremsstrahlung, and solid lines to the combined $v_2$ of photons from \twototwo and bremsstrahlung processes. Bands denote the statistical uncertainty.}
  \label{fig:v2_smash}
\end{figure}

This transition from predominantly bremsstrahlung photons to \twototwo photons also reflects itself in the composition of the $v_2$ coefficients shown in Fig.~\ref{fig:v2_smash}. For readability, we restrict the therein presented results to those extracted from SMASH and omit results from MUSIC, as the qualitative message is identical. Besides, the very same results are already depicted in Fig.~\ref{fig:v2spectra}, although arranged differently.
The \pt{} dependence of the photon elliptic flow coefficient $v_2$ is presented in Fig.~\ref{fig:v2_smash}, split into contributions from \twototwo scatterings, bremsstrahlung processes and their combined total. They are marked with dashed lines, dotted lines and solid lines, respectively. Results for RHIC are again presented on the left, those for LHC on the right. Regarding the composition of the total $v_2$ it becomes evident that at low \pt{} it is largely dominated by bremsstrahlung photons, while for rising \pt{} the influence of \twototwo photons increases until it is dominated by the $v_2$ of photons produced in \twototwo scatterings. This behaviour is perfectly in line with observations made in Fig.~\ref{fig:ratio_spectra}.
Therefore, we conclude that the bremsstrahlung photons from the hadronic stage are mainly responsible for the higher elliptic flow at low transverse momenta in the non-equilibrium scenario.

\subsection{\label{sec:v2}Time evolution of the $v_2$ coefficients}

In this section, the time-evolution of the photon elliptic flow, and its $\pt{}$ integrated version are analyzed,
considering only photons produced in the non-equilibrium afterburner. As both LHC and RHIC settings show qualitatively similar properties, in what follows, the results for RHIC will be presented. \\

\begin{figure}[t]
  \includegraphics[width=0.48\textwidth]{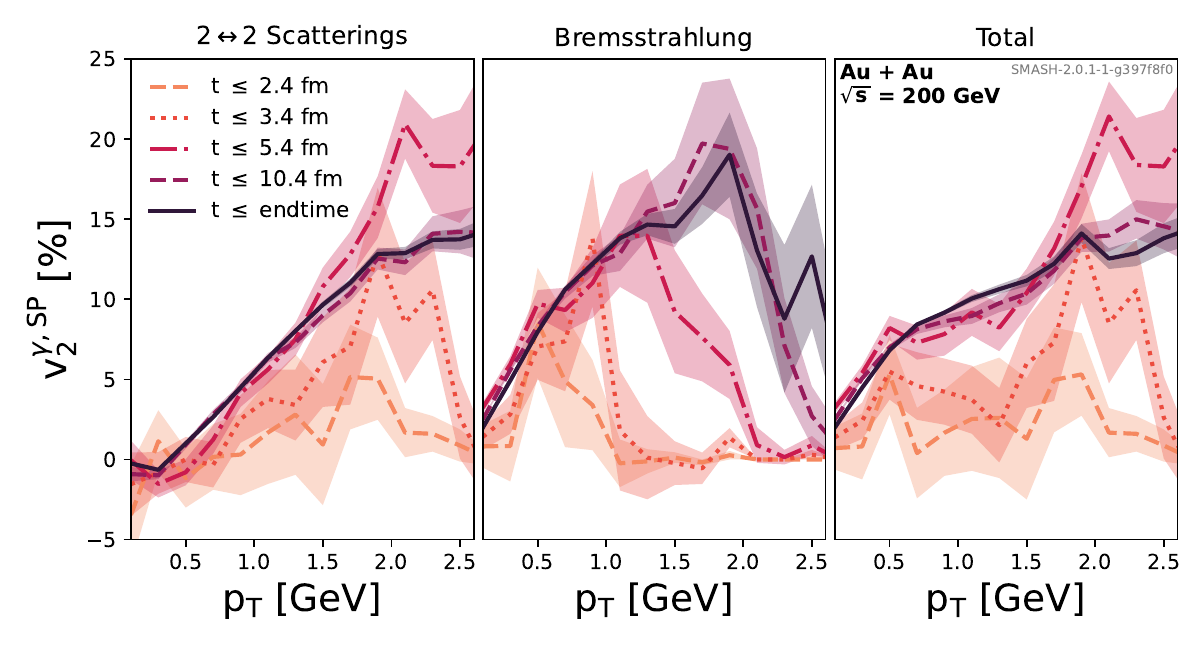}
  \caption{Time evolution of photon $v_2$ as computed in the hadronic afterburner (``setup A''). The left panel contains the $v_2$ of photons originating from \twototwo scatterings, the center panel the $v_2$ of photons from bremsstrahlung processes, and the right panel their weighted average. The
  different lines correspond to different times, up to which photons are considered to determine $v_2$. Bands denote the statistical uncertainty.}
  \label{fig:v2_evolution}
\end{figure}

\begin{figure}[t]
  \includegraphics[width=0.408\textwidth]{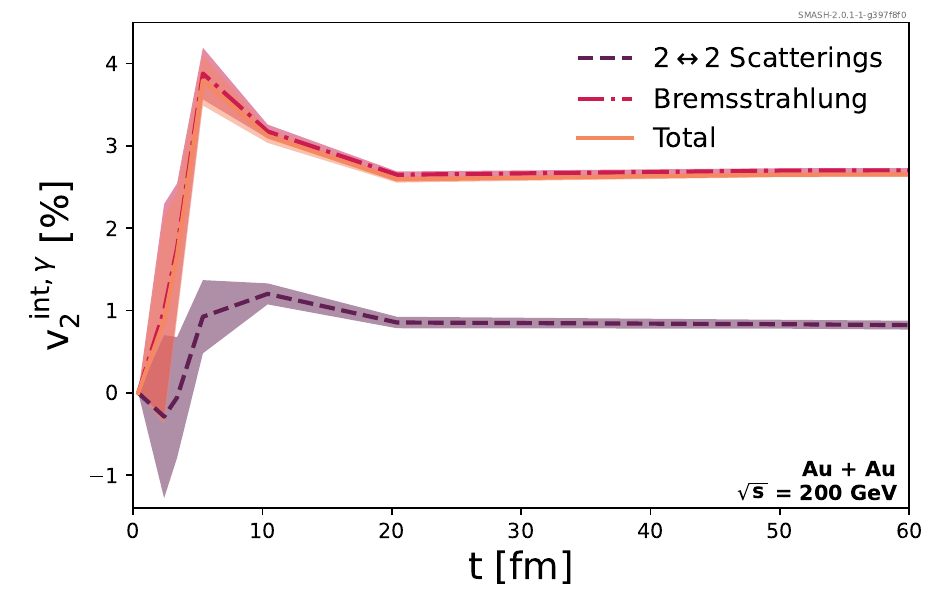}
  \caption{Integrated $v_2$ of photons as a function of time, split by their process origins and obtained with the SMASH afterburner. Photons from \twototwo scatterings are displayed with the dashed line, photons from bremsstrahlung processes with the dashed-dotted line and their combination with the solid line. Bands denote the statistical uncertainty.}
  \label{fig:int_v2}
\end{figure}

In Fig. \ref{fig:v2_evolution} the evolution of the photon anisotropies is displayed. The value presented as a function of $t$ is to be understood as comprising all photons produced before time $t$, where $t=0$ corresponds to the initial time of the simulation, when the nuclei start interacting.
For both hadronic scattering channels, such anisotropies rise rapidly in time and later descrease slightly.
This increase and subsequent decrease can also be observed in Fig.~\ref{fig:int_v2}, where the integrated $v_2$ is displayed as a function of time.
Therein, it further becomes apparent that the total integrated $v_2$ is vastly dominated by photons produced in bremsstrahlung processes,
which, in particular at low \pt{}, constitutes the dominant photon contribution (c.f. Fig.~\ref{fig:ratio_spectra}).
Furthermore, considering the entire \pt{} range, $ \approx 95$ \% of all photons stem from bremsstrahlung processes while only $\approx 5$ \% are produced in \twototwo scatterings.

The increase of photon $v_2$ in Figures~\ref{fig:v2_evolution} and ~\ref{fig:int_v2} can be attributed to the dynamical production of thermal pions (and other resonances) from the hypersurface. As time increases, more parent pions come to existence, and interact with each other, as well as with $\rho$ mesons. The resulting photons produced in these interactions are imprinted with the anisotropies carried from the transition hypersuface, and then converted by the sampler. These anisotropies originate directly from the hydrodynamical evolution at transition time and, thanks to the fact the hadronic transport is less effective at isotropization than hydrodynamics, the signal sees less suppression at late times. \\ \indent
On the other hand, the leading contribution to the decrease of this peak
can be attributed to a dynamical feed-down effect. It is a well known phenomenon (see Ref. \cite{Steinheimer:2017vju}) that the decay of higher-mass resonance states modifies the total anisotropies of pions. As the aforementioned resonances are, by their masses, less sensitive to flow, they will decay to pion states with less resulting $v_2$. The decrease of photon anisotropies in time is then a result of the depletion of high-mass resonances. In the presented setting this is a dynamical process. Since resonances appear along the freezeout hypersurface and their decays are delayed by boosts, the feed-down reduction of the anisotropies is then a time resolved process.

%
%

\section{\label{sec:Outlook}Conclusion and Outlook}
A hybrid approach consisting of ideal hydrodynamics (MUSIC) and hadronic transport (SMASH) is applied to study photon production in Au+Au collisions at $\sqrt{\mathrm{s}_\mathrm{NN}}$ = 200 GeV and Pb+Pb collisions at $\sqrt{\mathrm{s}_\mathrm{NN}}$ = 2.76 TeV.
We assessed the impact of non-equilibrium dynamics in the hadronic afterburner, compared to a pure hydrodynamical evolution. Identical photon production channels are considered in the non-equlibrium and the local equilibrium setup to provide a methodical comparison framework.
It is found that the non-equilibrium setup leads to a softer late-stage photon spectrum (more low \pt{} photons, less high \pt{} photons), as well as larger $v_2$ at low and intermediate \pt{}, when compared to photons estimated by combining ideal hydrodynamics with thermal photon emission rates. Similar results are found both at RHIC and the LHC.

Once combined with contributions from the hydrodynamics phase above T = 150 MeV, the differences found in \pt{} spectra are negligible. In the differential  $v_2$, however, a clear enhancement is found for photons characterized by \pt{} $\lesssim 1.4$ GeV. For higher \pt{} the combined elliptic flow is strongly dominated by photons produced above the switching temperature.

Although obtained in a simplified setup omitting event-by-event fluctuations and viscosities, the presented results demonstrate the importance of realistic non-equilibrium dynamics in the late stages of heavy-ion collisions in view of photon production. In particular at low \pt{} they play a crucial role for photon momentum anisotropies. To assess the implications of the late stage hadronic non-equilibrium dynamics on the direct photon puzzle and compare to experimental results in more detail, it will be important to repeat the presented study including event-by-event fluctuations and viscous corrections in addition to the inclusion of the ``primordial'' photons computed using pQCD \cite{Paquet:2015lta}. In doing so deviations from local equilibrium in the hydrodynamics phase above and below the switching temperature can be captured properly. It will however introduce additional challenges, in particular from the uncertainty in mapping viscous hydrodynamics to a hadronic momentum distribution that deviates from the thermal one.

Furthermore, only contributions from $\pi$-$\rho$ scatterings and $\pi$-$\pi$ bremsstrahlung are considered for the presented work. To fully describe hadronic photon production in its entirety it will be important to extend this framework by additional meson-baryon and baryon-baryon interactions. These efforts are however beyond the scope of this work, and are left for future work.

\begin{acknowledgements}
This project was supported by the DAAD funded by BMBF with Project-ID 57314610 as well as by the Deutsche Forschungsgemeinschaft (DFG, German Research Foundation) – Project number 315477589 – TRR 211. A.S. acknowledges support by the Stiftung Polytechnische Gesellschaft Frankfurt am Main as well as the GSI F\&E program. A.S. is further grateful to Markus Mayer for providing the flow analysis framework. J.-F.P. is supported in part by the U.S. Department of Energy Grant no. DE-FG02-05ER41367, and C.G. is supported in part by the Natural Sciences and Engineering Research Council of Canada. This work is part of a project that has received funding from the European Union’s Horizon 2020 research and innovation programme under grant agreement STRONG – 2020 - No 824093. Computational resources have been provided by the Center for Scientific Computing (CSC) at the Goethe-University of Frankfurt and the GreenCube at GSI. Computations were also made on the supercomputer B\'{e}luga, managed by Calcul Qu\'{e}bec and Compute Canada. The operation of this supercomputer is funded by the Canada Foundation for Innovation (CFI), Minist\`{e}re de l'\'{e}conomie et de l'Innovation du Qu\'{e}bec (MEI) and le Fonds de recherche du Qu\'{e}bec (FRQ).
\end{acknowledgements}


\appendix

\section{\label{app:SMASH_brems}Bremsstrahlung in SMASH}
\begin{figure}[t]
  \centering
  \includegraphics[width=0.45\textwidth]{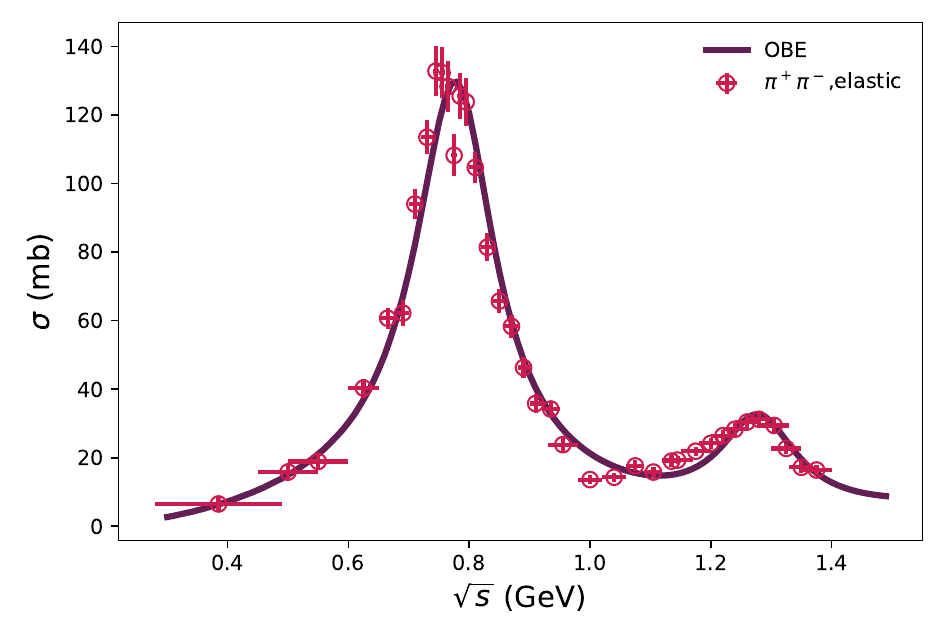}
  \caption{The $\pi^+\pi^-$ elastic cross-section. Experimental data from refs.\cite{Srinivasan:1975tj,Protopopescu:1973sh} was used to fit the coupling parameters. }
  \label{fig:pi_pi_elastic}
  \end{figure}

  In this work, photon bremsstrahlung from meson scattering has been introduced into SMASH, completing the leading contributions for photon production in a hadronic transport approach. In this appendix, a brief summary is presented for the theoretical input used in the computation of the photon radiation from pions, starting from the underlying model. Pion elastic scattering has been described successfully using the One Boson Exchange (OBE) model, derived in the context of chiral perturbation theory \cite{Ogawa:1967iv,Li:2002yd}. In this model, $\pi$ mesons are taken to be stable and interact by the exchange of unstable scalar $\sigma$, vector $\rho$, or tensor $f_2(1270)$ resonances. The Lagrangian for the OBE model consists of kinetic parts for the stable and unstable fields, as well as an interaction term,
  \begin{align}
    \begin{split}
      \mathcal{L}_{\mathrm{int}} \ = \ &\gs\, \sigma \,\partial_\mu\pi_a \,\partial^\mu \pi_a \ + \ \gr\, \epsilon_{abc}\,\rho^{\mu}_a\, \pi_b \,\partial_\mu\pi_c \ + \ \\ &\gf\, f_{\mu\nu} \,\partial^\mu\pi_a \,\partial^\nu \pi_a
      \label{eq:OBELagrangian}
    \end{split}
  \end{align}
  where $a,b,c={1,2,3}$ are the component indices of the $\pi$ and $\rho$ iso-triplets. The kinetic parameters of the model, that is the masses and resonance widths, are fixed to match the SMASH degrees of freedom. For the pions, $m_\pi=0.138 \GeV$ is used for the charged as well as the neutral pions. For the resonances, we use $\ms=0.8 \GeV$, $\GS=0.52 \GeV$,  $\mr=0.776\GeV$, $\GR=0.149\GeV$ , $\mf=1.276\GeV$ and $\GF=0.16\GeV$. The couplings, on the other hand, have been fitted to the experimental data, which can be seen in Fig. \ref{fig:pi_pi_elastic}.
  Their values are determined to be $\gs=5.377\GeV^{-1}$, $\gr=6.015$ and $\gf=4.33\GeV^{-1}$. As it has been extensively done in the literature \cite{Linnyk:2015tha,Eggers:1995jq}, also in this work, the finite size of resonances is effectively accounted for by suppressing high-momentum transfers. This is done by including the following form factors in the $u$- and $t$-channels,
  \begin{equation}
      h_\alpha(p^2) = \frac{m^2_\alpha-m_\pi^2}{m^2_\alpha-p^2},
  \end{equation}
  where $\alpha=\{\sigma,\rho,f\}$.

  Electromagnetic interactions can be built into the Lagrangian (eq. \ref{eq:OBELagrangian}) by upgrading it to a local $U(1)$ theory. This creates new interaction operators,
  \begin{equation}
      \mathcal{L}_{\mathrm{int}}=\mathcal{L}_{\pi\pi\gamma}+\mathcal{L}_{\rho\rho\gamma}+\mathcal{L}_{\pi\pi\sigma\gamma}+\mathcal{L}_{\pi\pi\rho\gamma}+\mathcal{L}_{\pi\pi f\gamma}
      \label{eq:OBELagrangianEM}
  \end{equation}
  which can be found explicitly in terms of the fields, as well as their respective Feynman rules, in Ref. \cite{Eggers:1995jq}. This new Lagrangian is used to compute the following channels,
  \vspace{+0.1cm}
  \begin{equation}
    \label{eq:photon_channels}
  \begin{split}
  &\pi^{+}\pi^- \rightarrow (\sigma,\rho^0,f)\rightarrow \pi^{+}\pi^-\gamma \\
  &\pi^{\pm}\pi^\pm \rightarrow (\sigma,\rho^0,f)\rightarrow \pi^{\pm}\pi^\pm\gamma\\
  &\pi^{\pm}\pi^0 \rightarrow \rho^\pm\rightarrow \pi^{\pm}\pi^0\gamma\\
  &\pi^{+}\pi^- \rightarrow (\sigma,\rho^0,f)\rightarrow \pi^{0}\pi^0\gamma \\
  &\pi^{0}\pi^0 \rightarrow (\sigma,\rho^0,f)\rightarrow \pi^{+}\pi^-\gamma
  \end{split}
  \end{equation}
  \vspace{+0.1cm}
  It is important to emphasize that by using this prescription, gauge symmetry is trivially built in, and the Ward-Takahashi identities are satisfied by any process one would like to compute. However, because of the inclusion of form factors for the resonance-$\pi$ vertices, gauge symmetry is effectively broken for the $\pi^{\pm}\pi^0$ channel. This happens due to the appearance of the $\mathcal{L}_{\rho\rho\gamma}$ term in the interaction Lagrangian, which means that a photon can be emitted by the charged $\rho$. This can be solved by introducing a new form factor in the internal line photon diagram and solving for it by enforcing the Ward-Takahashi identity, $k_\mu\mathcal{M}^\mu=0$.

\begin{figure}[t]
  \centering
  \includegraphics[width=0.45\textwidth]{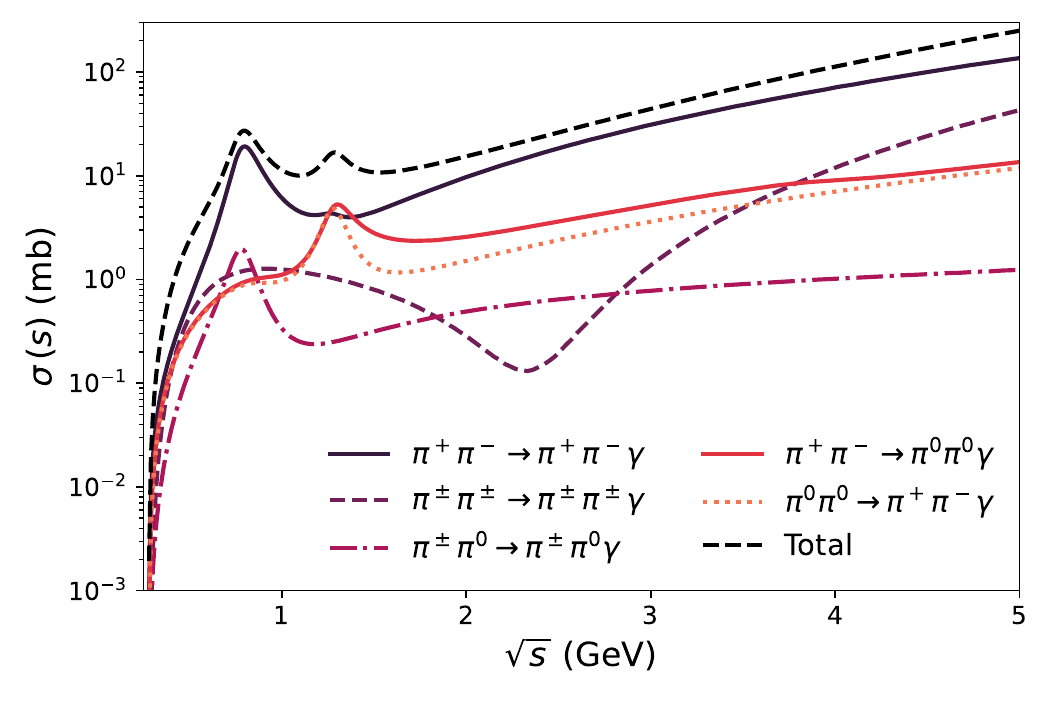}
  \caption{The total cross sections for the seven channels of photon bremsstrahlung in pion-pion scatterings for photons with an energy above $k = 0.001\GeV$.}
  \label{fig:photon_cs_tot}
  \end{figure}

  Once the amplitudes are all accounted for, the differential photon cross-sections can be computed for the seven channels using the standard formulae for 2 $\to$ 3 scatterings \cite{Byckling:1971vca}. The total cross-sections (see Fig. \ref{fig:photon_cs_tot}), as well as the single differential cross-section for both the magnitude photon momentum $k$, and the angle $\theta_k$ with respect to the momentum axis in the rest frame of the incoming pion pair are used as input for SMASH. The cross-sections which involve photon momentum integrations are obtained by the integration of $\rmd\sigma/\rmd k$ down to a cut-off of $k=0.001\GeV$.\\
  The cross sections used for this work are embedded in the PHOXTROT project and thus publicly available on GitHub at \url{https://github.com/smash-transport/phoxtrot}.\\
  \newline
  The photon production in SMASH relies on a perturbative treatment, motivated by the fact that photons are unlikely to interact with the hadronic medium post production.
  Hence, a bremsstrahlung process is performed whenever the initial state particles of a hadronic scattering correspond to any of the processes listed in \ref{eq:photon_channels}. This process and the produced photon are directly printed to the output, but the final state particles are not propagated further in the evolution. Instead, the underlying hadronic interaction is performed as if no photon process had occurred. The produced photon needs then be assigned a specific weight to account for the reduced probability of a photon process as compared to a hadron interaction. This weight is defined as:
  \begin{equation}
    W = \frac{\sqrt{\frac{\mathsf{d} \sigma_\gamma}{\mathrm{d} k} \Delta k \  \frac{\mathsf{d} \sigma_\gamma}{\mathrm{d} \theta_k} \Delta \theta_k }}{\mathsf{N_{frac}} \ \sigma_\mathsf{had}}
    \label{eq:weight}
  \end{equation}
  Here, $\frac{\mathsf{d} \sigma_\gamma}{\mathrm{d} k}$ and $\frac{\mathsf{d} \sigma_\gamma}{\mathrm{d} \theta_k}$ are the differential cross sections with respect to $k$ and $\theta_k$, and $\Delta k$ and $\Delta \theta_k$ the ranges in which the kinematic properties of the produced photons are sampled. $N_\mathrm{frac}$ denotes the number of fractional photons and $\sigma_\mathsf{had}$ the hadronic cross section of the underlying interaction.
  Fractional photons are a mean to properly sample the angular distributions. For each underlying hadronic interaction $N_\mathrm{frac}$ fractional photons are sampled with different kinematic properties provided by the ranges for $k$ and $\theta_k$:
  \begin{align}
    k \in [0.001, \dfrac{s - 4 \ m_\pi^2}{2 \ \sqrt{s}}] \qquad \qquad \theta_k \in [0, \pi]
  \end{align}
  In the above, $m_\pi$ is the pion mass and $\sqrt{s}$ the center-of-mass energy of the incoming pions.\\
  \begin{figure}[t]
    \centering
    \includegraphics[width=0.45\textwidth]{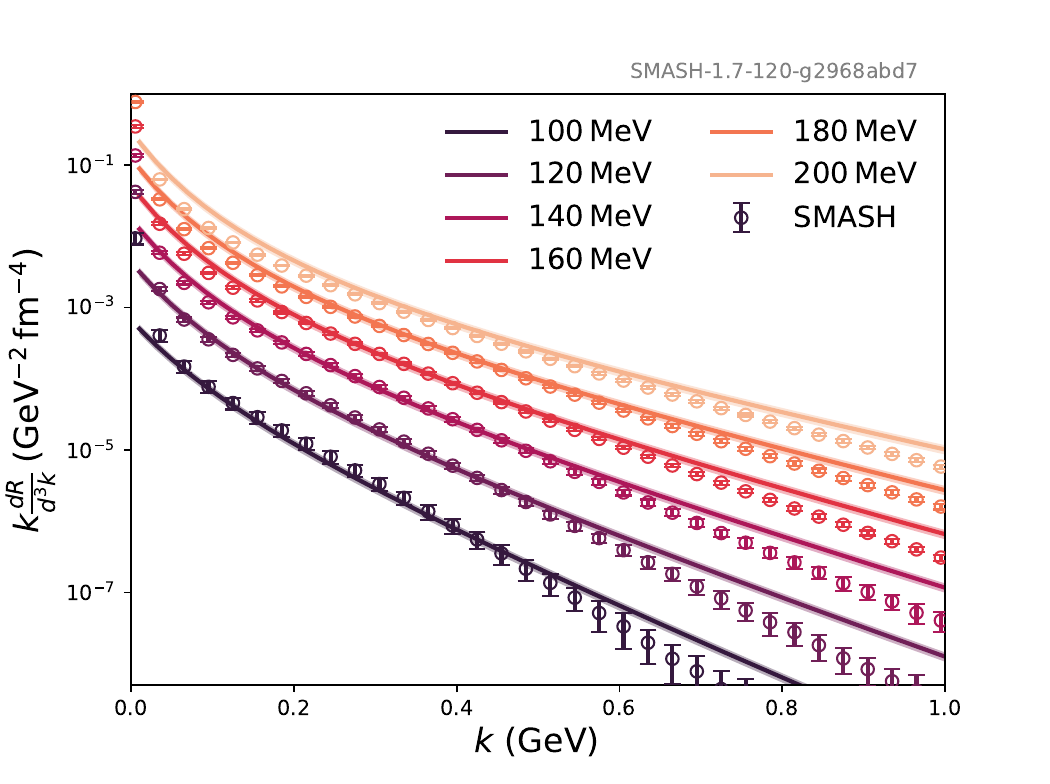}
    \caption{Comparison of the total thermal bremsstrahlung rate as computed in an infinite matter set-up with SMASH (points), to parametrizations taken from \cite{Heffernan:2014mla} (lines). The deviations at higher photon energies stem from inconsistencies in the kinetic parameters of the OBE model underlying the cross sections and rates.}
    \label{fig:brems_rates_comp}
  \end{figure}
  To validate the implementation in SMASH, an infinite matter simulation is applied to extract the thermal equilibrium photon rates corresponding to the processes listed in \ref{eq:photon_channels}. For this, a cubic box with periodic boundary conditions, filled with $\pi$ and $\rho$ mesons, is thermally initialized at different temperatures between T = 100 MeV and T = 200 MeV and evolved for t = 100 fm. The resulting thermal photon rates, summing all contributions from the processes listed in \ref{eq:photon_channels} are displayed in Fig.~\ref{fig:brems_rates_comp} and compared to parametrizations of these rates, taken from~\cite{Heffernan:2014mla}.
  It can be seen that a nearly-perfect agreement is obtained in the low and intermediate photon energy regime. Above $k \approx$ 0.5 GeV however, the SMASH results drop below the rates provided by the parametrization. This is due to inconsistencies between the kinetic parameters of the OBE model used to determine the rates in Ref.~\cite{Liu:2007zzw}, underlying the parametrizations, and those used to determine the cross sections for SMASH. For the latter, the kinetic parameters are chosen such that they match the SMASH degrees of freedom. Small differences in the resulting photon rates, that become more significant for higher photon energies, are therefore expected. Still, the good agreement in the low and intermediate energy regime serves as a validation of the bremsstrahlung cross sections as well as their implementation in SMASH.
  These results could further be systematically improved by extension to bremsstrahlung processes involving higher mass resonances, as has been done within the soft photon approximation in \cite{Linnyk:2015tha}.

\bibliography{Afterburner_Photons}

\end{document}